\documentclass[twocolumn, aps,prl,superscriptaddress,showpacs,preprintnumbers,amsmath,amssymb]{revtex4-1}
\usepackage{graphicx,color}
\usepackage{dcolumn}
\usepackage{colordvi} 
\usepackage{bm}

\usepackage{amsmath,amssymb,amsfonts}
\usepackage[]{color}

\begin{document}
\title{Gravitational Wave Memory: A New Approach to Study Modified Gravity}
\author{Song Ming Du}
\email{smdu@caltech.edu}
\affiliation{Theoretical Astrophysics 350-17, California Institute of Technology, Pasadena, California 91125, USA}

\author{Atsushi Nishizawa}
\affiliation{Theoretical Astrophysics 350-17, California Institute of Technology, Pasadena, California 91125, USA}
\affiliation{Department of Physics and Astronomy, The University of Mississippi, University, Mississippi 38677, USA}

\date{\today}

\begin{abstract}
It is well known that two types of gravitational wave memory exist in general relativity (GR): the linear memory and the non-linear, or Christodoulou memory. These effects, especially the latter, depend on the specific form of Einstein equation. 
It can then be speculated that in modified theories of gravity, the memory can differ from the GR prediction, and provides novel phenomena to study these theories. We support this speculation by considering scalar-tensor theories, for which we find two new types of memory: the T memory and the S memory, which contribute to the tensor and scalar components of gravitational wave, respectively.  In particular, the former is  caused by the burst of energy carried away by scalar radiation, while the  latter is intimately related to the no scalar hair property of black holes in scalar-tensor gravity. We estimate the size of these two types of memory in gravitational collapses, and formulate a detection strategy for the S memory, which can be singled out from tensor gravitational waves. We show that (i) the S memory exists even in spherical symmetry, and is observable under current model constraints, and (ii) while the T memory is usually much weaker than the S memory, it can become comparable in the case of spontaneous scalarization. 
\end{abstract}

\maketitle

\noindent {\it Introduction.--} The discovery of GW150914 \cite{Abbott} by advanced LIGO marks the beginning of a new era in gravitational physics, and brings forth new opportunities to study properties of black holes and to test theories of gravity. In this letter, we will show that both objectives can be met using gravitational wave memory.

The gravitational wave memory is a permanent change in spacetime geometry, which in general relativity (GR) is a jump in the transverse-traceless part of the spacetime metric $\Delta h^{\text{TT}}_{ij}$ before and after a burst event \cite{Braginsky}. Gravitational-wave memory was first predicted in the 1970s as originating from an overall change in the source term of the linearized Einstein equation \cite{Zel'dovich,Smarr,Bontz,Kovacs,Braginsky2}.  This is now referred to as the ``linear memory''.  Decades later, Christodoulou found that nonlinearities of the Einstein's equation lead to another memory~\cite{Christodoulous}, which is now referred to as the {\it nonlinear} or {\it Christodoulou memory}.  Shortly after, the nonlinear memory was interpreted as sourced by bursts of gravitational radiation \cite{Thorne2, Wiseman}.

Since both memories depend on the specific form of the field equation, one can speculate that other types of memory may arise in modified theories of gravity.  Modifications to GR seems inevitable if one considers general relativity as a low-energy effective theory to a quantum theory of gravity, which adds new terms to the Einstein-Hilbert action, such as higher-order curvature terms, or extra scalar degree of freedom coupled to the tensor degrees of freedom~\cite{Clifton, Berti}. In this letter, we carry out a proof-of-principle discussion for new memory effects in scalar-tensor theories of gravity. Using a perturbative treatment, we will first show that a T memory arises in the tensor components of gravitational wave due to energy carried away by scalar radiation.  We will then point out that the no scalar hair property of black holes in scalar-tensor theories will give rise to the S memory, a scalar component of gravitational wave.   We will go on to estimate the size of both memories using the simplest progenitor model.     Finally, we will formulate a detection strategy that targets the S memory, and consider detectability using the current and next generations of ground-based gravitational wave detectors.

\noindent {\it T memory and S memory in scalar-tensor gravity. --} Scalar-tensor theories are a simple but attractive class of modified theory, e.g., they can be viewed as arising from dimensional compactification of higher dimensional theories~\cite{Fuji}.  Let us consider a single scalar field $\phi$ and a $\phi$-dependent coupling constant $\omega(\phi)$, with an action of
\begin{align}
&S=\int d^4 x \ \left\{\sqrt{-g}\ \left(\phi R -\frac{\omega(\phi)}{\phi}\partial^\mu\phi\partial_\mu\phi \right) +16\pi\mathcal{L}_{\text{M}}\right\},\nonumber
\end{align} 
where $R$ is the Ricci scalar associated with the spacetime metric $g_{\mu\nu}$, $\mathcal{L}_{\text M}$ is the matter-sector Lagrangian which depends on $g_{\mu\nu}$ and matter fields. We start  by expanding the metric and the scalar field as $g_{\mu\nu}=\eta_{\mu\nu}+h_{\mu\nu}$ and $\phi=\phi_0+\delta\phi$, where $\eta_{\mu\nu}=\text{diag}(-1,1,1,1)$ and $\phi_0$ is the value of the scalar field at null infinity, which is related to Newton's constant via $G\phi_0=(2\omega(\phi_0)+4)/(2\omega(\phi_0)+3)$. The action is then expanded as:
\begin{align}
S= \int d^4 x\ \left\{ \mathcal{L}_{\text{ST}}^0+\mathcal{L}_{\text{ST}}^1 + 16\pi\mathcal{L}_{\text{M}}^0+16\pi\mathcal{L}_{\text{M}}^1 +...\right\}. \nonumber
\end{align}
Here $\mathcal{L}_{\text M}^0$ is the matter Lagranian in flat spacetime and 
\begin{align}
&\mathcal{L}_{\text {ST}}^0=\frac{\phi_0}{2} H^{\mu\nu}\mathcal{V}_{\mu\nu\rho\sigma}H^{\rho\sigma}-\frac{\alpha_0^{-2}}{2\phi_0}\partial^\mu\delta\phi \partial_\mu \delta\phi \label{3}\,,\\
&\mathcal{L}_{\text {M}}^1=\frac{1}{2} H^{\mu\nu} T_{\mu\nu}-\frac{1}{2}\frac{\delta\phi}{\phi_0}T\label{4}\,,\\
&\mathcal{L}_{\text {ST}}^1=\frac{\alpha_0^{-2}}{2\phi_0}\left(H^{\mu\nu}-\frac{1}{2}\eta^{\mu\nu} H\right) \partial_\mu\delta\phi\partial_\nu\delta\phi\nonumber\\
&\qquad\quad-\frac{\alpha_0^{-4} \beta_0}{2\phi_0^2}\delta\phi\, \partial^\mu \delta\phi \partial_\mu\delta\phi    \label{5}\, ,
\end{align}
where $\mathcal{L}^0_{\text{ST}}$ and $\mathcal{L}^1_{\text{M}}$ are kinetic and source terms of $h_{\mu\nu}$ and $\delta\phi$ respectively, while we kept the leading coupling term between them and the leading self-interactive term of $\delta\phi$ in $\mathcal{L}^1_{\text{ST}}$. Here all indices are raised and lowered by $\eta_{\mu\nu}$, and $T_{\mu\nu}$ is the stress-energy tensor of matter. In order to eliminate the kinetic term crossing between $h_{\mu\nu}$ and $\delta\phi$ in the original expansion, we redefined the physical degrees of freedom as $H_{\mu\nu}=h_{\mu\nu}+\eta_{\mu\nu}\phi_0^{-1}\delta\phi$. 
The operator $\mathcal{V}_{\mu\nu\rho\sigma}$ is given by 
\begin{align}
2 \mathcal{V}_{\mu\nu\rho\sigma}=&(\eta_{\mu\rho}\eta_{\nu\sigma}-\eta_{\mu\nu}\eta_{\rho\sigma})\partial^2+ \eta_{\mu\nu}\partial_\rho\partial_\sigma +\eta_{\rho\sigma}\partial_\mu\partial_\nu\nonumber\\
& -\eta_{\mu\rho}\partial_\nu\partial_\sigma -\eta_{\nu\sigma}\partial_\mu\partial_\rho.\nonumber
\end{align}
Up to leading order, we expand $\omega(\phi)$ as $\omega(\phi_0)+\omega'(\phi_0)\delta\phi$ and adopt the often used parameters: $\alpha_0 = (2\omega(\phi_0)+3)^{-\frac{1}{2}}$ and $\beta_0 = 2\phi_0\omega'(\phi_0)/(2\omega(\phi_0)+3)^{2}$. Note that all terms in Eqs. (\ref{3})--(\ref{5}) are invariant under infinitesimal diffeomorphisms: $H'_{\mu\nu}=H_{\mu\nu}-\partial_\mu\xi_\nu-\partial_\nu\xi_\mu$ and $\delta\phi'=\delta\phi$.

The first term of Eq. (\ref{3}) gives the vacuum field equation for $H_{\mu\nu}$: $\mathcal{V}_{\mu\nu\rho\sigma}H^{\rho\sigma}=0$, the same as in GR. We can then similarly take the Lorenz gauge $\partial^\mu H_{\mu\nu}-\partial_\nu H/2=0$ and  use infinitesimal diffeomorphisms to gauge away redundant degrees of freedom.  In this way, only two physical degrees of freedom are left for $H_{\mu\nu}$. However, gravitational-wave detectors are sensitive to $h_{\mu\nu}$, which depends on both $H_{\mu\nu}$ and  $\delta\phi$.  We can further gauge away remaining non-physical degrees of freedom which leaves $h_{ij}=h^{\text{T}}_{ij}+h^{\text{S}}_{ij}$, where
\begin{align}
h^{\text{T}}_{ij}=h_+ e^+_{ij}+h_\times e^\times_{ij},\qquad h^{\text{S}}_{ij}=h_\circ e^\circ_{ij}. \label{6}
\end{align}
Here the polarization tensors are defined by $e^+_{ij}=\hat{m}_i\hat{m}_j-\hat{n}_i\hat{n}_j$, $e^\times_{ij}=\hat{m}_i\hat{n}_j+\hat{n}_i\hat{m}_j$ and $e^\circ_{ij}=\hat{m}_i\hat{m}_j+\hat{n}_i\hat{n}_j$, where $\hat m$, $\hat n$ are spatial unit vectors orthogonal to the  wave propagation direction $\hat\Omega$. The amplitudes are related to the tensor $H_{\mu\nu}$ and the scalar $\delta \phi$ via
\begin{align}
\!\!h_+=e_+^{ij}H^{\text{TT}}_{ij}/2,\; h_\times=e_\times^{ij}H^{\text{TT}}_{ij}/2,\; h_\circ = -\phi_0^{-1}{\delta\phi},\label{7}
\end{align}
where $H_{ij}^\text{TT}$ denotes the transverse-traceless part of $H_{ij}$ \cite{Misner}. We shall refer to the  $+$, $\times$ modes as tensor (T) modes, and the $\circ$ mode as the scalar (S) mode.

The first term of Eq. (\ref{4}) indicates that $H_{\mu\nu}$ is sourced by the stress-energy tensor $T_{\mu\nu}$.  In addition, coupling between $H_{\mu\nu}$ and the quadratic terms of $\delta\phi$ in Eq. (\ref{5}) provides another source for $H_{\mu\nu}$. In fact, we can define
\begin{align}
8\pi\mathcal{T}_{\mu\nu}&=\frac{\partial \mathcal{L}^1_{\text{ST}}}{\partial H^{\mu\nu}}=\frac{\omega(\phi_0)}{\phi_0}\left(\partial_\mu\delta\phi\partial_\nu\delta\phi-\frac{\eta_{\mu\nu}}{2}\partial^\rho\delta\phi\partial_\rho\delta\phi\right) \label{8}
\end{align}
as the effective stress-energy tensor of the scalar radiation. Just as the Christodoulou memory is caused by the burst of {\it gravitational} radiation, we expect the burst of {\it scalar} radiation would generate a new gravitational wave memory, which we call the T memory in scalar-tensor gravity since it contributes to the tensor components of gravitational wave. Following a similar argument as \cite{Wiseman}, the T memory can be expressed as
\begin{equation}
\Delta h^{\text T}_{ij} =\frac{4}{\phi_0 r}
\int d\hat\Omega'\int dt\ r^2 \mathcal{T}^{0k}\hat\Omega'^k
\left(\frac{ \hat\Omega'^i \hat\Omega'^j}{1-{\hat \Omega'}\cdot {\hat\Omega}}\right)^{\text{TT}}.\label{9}
\end{equation}
Here the spatial vector ${\hat\Omega}$ is the wave-propagation direction, ${\hat\Omega'}$ is a unit  vector integrated over all sky directions, and $\mathcal{T}^{0k}$ is the effective energy flux of scalar radiation. Note that the T memory vanishes in spherical symmetry.  \footnotetext{Recently the extra memory effect due to the radiation of scalar energy is discussed in the circumstance of compact binary systems in \cite{Lang}, which is equivalent to our T memory.}
  
Let us turn now to the scalar degree of freedom. From the second term in Eq.(\ref{4}), the scalar field $\delta\phi$ is sourced by the trace of stress-energy tensor of matter. This means any cold matter ($p\ll \rho$) can change the scalar field from its value at null infinity $\phi_0$. We shall refer to the resulting $\phi_{\text H}$ inside and outside a star as its  scalar field profile. However, in 1972 Hawking discovered that black holes in Brans-Dicke theory are the same as in GR: they have no scalar hair and $\phi_{\text H} = \phi_0$ everywhere~\cite{Hawking}.  This was also shown to be true for general scalar tensor theories~\cite{Sotiriou}. The no scalar hair theorem has the following consequence: in any gravitational collapses resulting in black holes, the scalar field outside the collapsing star changes from $\phi_{\text H}$ to $\phi_0$, and 
~(\ref{6}) and (\ref{7}) this causes a permanent change in the scalar component $h^S_{ij}$: 
\begin{align}
\Delta h^{\text{S}}_{ij}= \phi_0^{-1}(\phi_{\text{H}}-\phi_0)\ e^\circ_{ij} \label{10}.
\end{align}
We shall call this the {\it S memory }.  Differently from other memories, the S memory: (i) exists even in spherical symmetry, and (ii) has a reverse temporal feature --- it begins with a non-zero initial value, and drops down to zero.

\begin{figure}[t]
\centering
\includegraphics[width=87mm]{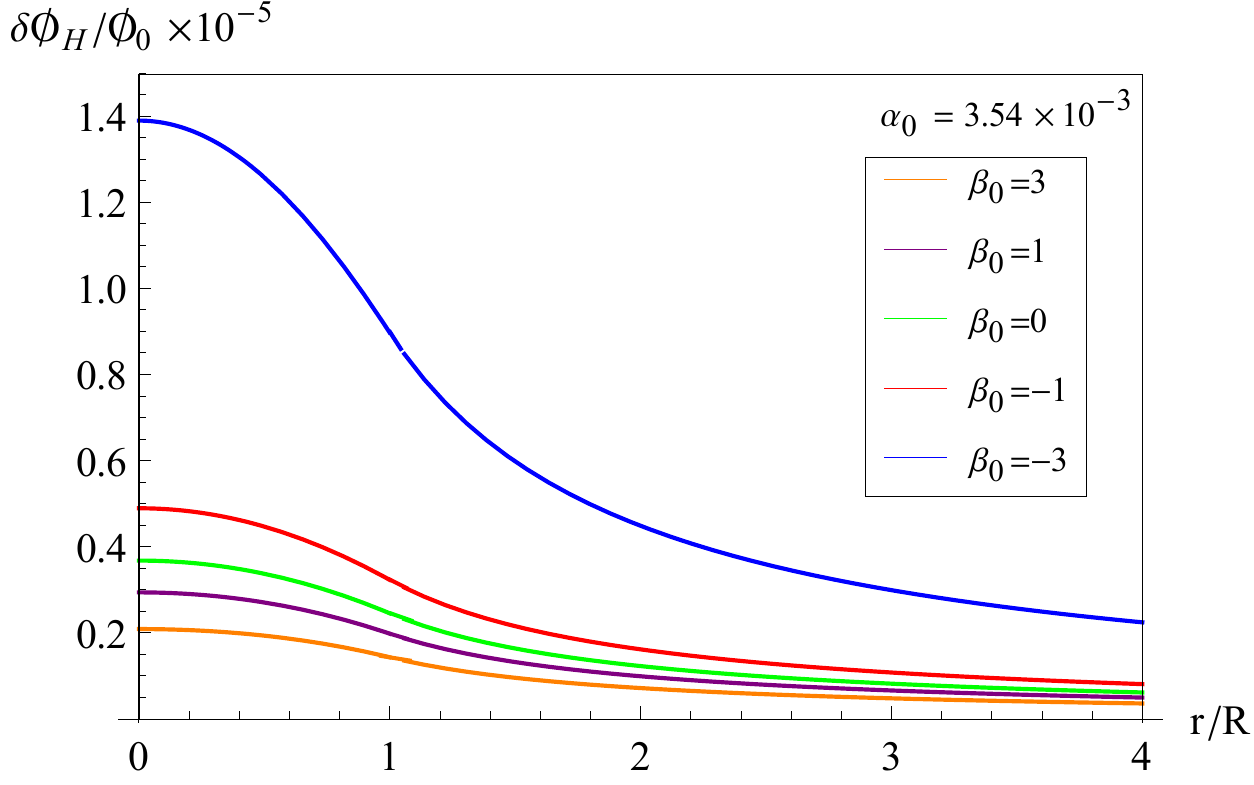}
\caption{The stationary interior and exterior scalar field profile of an $M=10M_\odot$ and $R=100M_\odot$ Newtonian star for different values of $\beta_0$.
}
\label{fig1}
\end{figure}

{\noindent {\it Analytic results of a simplified model.--}}
To estimate the size of the S and T memories, let us analyze a spherically symmetric and homogeneous Newtonian star $(p\ll\rho)$,  by solving the linearized field equation for $\delta \phi$ obtained from Eqs. (\ref{3})--(\ref{5}), which reads:
\begin{align}
\partial^2 \delta\phi=8\pi (\alpha_0^2 -\phi_0^{-1}\beta_0\delta\phi) T. \label{11}
\end{align}
Here the trace of stress-energy tensor $T$ is $-3M/4\pi R^3$ inside the star and 0 outside the star, where $M$ and $R$ are the mass and radius of the star respectively. In this equation we dropped the non-linear terms which are lowered by a factor of $M(\phi_0 R)^{-1}$. By taking the continuity condition for the scalar field profile and its first-order derivative at the boundary $(r=R)$, the asymptotic condition $\phi_{\text H}({\infty})=\phi_0$ at the null infinity and requiring that there is no singularity inside the star, the stationary interior scalar field profile $\delta\phi_\text{H}(r)=\phi_{\text{H}}(r)-\phi_0$ is given by
\begin{align}
\delta\phi_{\text{H}}(r<R)=\frac{\phi_0 \alpha_0^2}{|\beta_0|}\times 
\begin{cases}
1-\frac{\sinh{\kappa r}}{\kappa r \cosh{\kappa R}}   \quad\beta_0>0\\
\frac{\sin{\kappa r}}{\kappa r \cos{\kappa R}}-1  \  \quad\beta_0<0
\end{cases}
\end{align}
while the stationary external scalar field profile is 
\begin{align}
\delta\phi_{\text{H}}(r>R)=\frac{\phi_0 \alpha_0^2}{|\beta_0|}\frac{R}{r}\times 
\begin{cases}
1-\frac{\tanh{\kappa R}}{\kappa R}   \quad\beta_0>0\\
\frac{\tan{\kappa R}}{\kappa R}-1   \ \quad\beta_0<0
\end{cases} \label{13}
\end{align}
where $\kappa\equiv(6M|\beta_0|\phi_0^{-1} R^{-3})^{1/2}$. Next we consider a progenitor with $M=10M_\odot$ and $R=100M_\odot$ as in \cite{Harada}. We plot the scalar field profile in Fig.~\ref{fig1} for different values of $\beta_0$ while saturate $\alpha_0$ to the Cassini bound \cite{Berti}. As we can see from the figure, the scalar field profile is amplified for negative values of $\beta_0$ and is depressed for positive $\beta_0$. This feature agrees with the results from numerical 
simulations \cite{Harada, Novak, Gerosa}.

From Eqs.~(\ref{10}) and (\ref{13}),  the S memory is given by
\begin{align}
\Delta h^{\text S}_{ij}=  \mathcal{N}(\beta_0,\mu)[{2\alpha_0^2 M}/({\phi_0 r})]e^\circ_{ij}.\label{14}
\end{align}
Here $\mu\equiv M/\phi_0 R$ and the $\beta_0$-scale factor $\mathcal{N}(\beta_0,\mu)$ is defined as:
\begin{align}
\mathcal{N}(\beta_0,\mu)=\frac{1}{2|\beta_0|\mu}\cdot
\begin{cases}
1-\frac{\tanh{\sqrt{6\mu|\beta_0|}}}{\sqrt{6\mu|\beta_0|}} \quad  \beta_0>0\\
\frac{\tan{\sqrt{6\mu|\beta_0|}}}{\sqrt{6\mu|\beta_0|}}-1 \ \quad \beta_0<0
\end{cases}.\nonumber
\end{align}
For the Brans-Dicke \cite{Brans} limit $\beta_0=0$, $\mathcal{N}(0,\mu)=1$. We also find that $\mathcal{N}(\beta_0,\mu)$ is singular at $\beta_0^{\text{crit}}=-\pi^2/(24\mu)$, which is $-4.11$ for the Newtonian star in Fig.~\ref{fig1}. For $\beta_0<\beta_0^{\text{crit}}$, solving Eq.~(\ref{11}) for a time dependent scalar field profile $\delta\phi_{\text H}(t,r)=f(r)e^{-i\omega t}$ will give a quasinormal mode $\omega_1$ with a negative imaginary part, which means the solution is not stable under small perturbations \footnotetext{Some negative values of $\beta_0$ are also inconsistent with the cosmological evolution of the scalar field and Solar System experiments, as explained in Appendix A of \cite{Sampson}.}. Physically, this corresponds to the effect called ``spontaneous scalarization", which was first discovered by Damour and Esposito-Farese \cite{Damour}.  For this reason, for $\beta<\beta_0^{\rm crit}$, we should use the fully nonlinear field equation instead of our leading-order approximation. Previous numerical simulations \cite{Novak2, Harada2} indicate that scalarization changes the asymptotic value of the scalar field profile from $\delta\phi_{\text H}\sim \alpha_0^2 M/r$ to $\delta\phi_{\text H}\sim \alpha_0 M/r$. Thus for $\alpha_0\sim 10^{-3}$, the scalar field profile for a scalarized star is enhanced by about 3 orders of magnitude.

Another important parameter is the time $\tau$ it takes $h^{\text S}_{ij}$ to change from $\Delta h^{\text{S}}_{ij}$ to zero. In our case, $\tau$ is the time for the progenitor collapse into a black hole \footnotetext{The time a star collapsing into a black hole is infinite long for an exterior observer, but here we use the {\it effective} collapse time, which is the 
time a star collapsing into its light ring as observed at null infinity.}. The gravitational collapse process for homogeneous spherical dust is described by Oppenheimer-Snyder model \cite{Oppenheimer}, which gives $\tau\simeq \pi R [(8GM/R)(1-2GM/R)]^{-\frac{1}{2}}$.
For $M=10M_\odot$ and $R=100M_\odot$, $\tau=1.93 \text{ms}$, the cut-off frequency of the memory \cite{Thorne} is $f_{\text{c}}=1/\tau\simeq 500 {\text{Hz}}$. This means that although the exact waveforms of scalar radiation in gravitational collapses have been studied from numerical simulations \cite{Shibata, Harada, Novak, Gerosa}, for ground based gravitational wave detectors, most of the detection band is dominated over by the memory as the ``zero-frequency limit" \cite{Smarr, Bontz}.

The T memory for compact binary systems has been discussed in \cite{Lang}, here we consider T memory in gravitational collapses and compare it with S memory.
From Eq.~(\ref{8}), a burst of energy flux carried by scalar field is generated when the scalar hair is radiated away in a short duration $\tau$, which means both S memory and T memory appear in gravitational collapses. The amplitude of T memory can be estimated from Eq.~(\ref{9}):
\begin{align}
\Delta h^{\text T}&\simeq\frac{4\epsilon}{\phi_0 r}\int\limits^{+\infty}_{-\infty}dt\ r^2\mathcal{T}^{0k}\hat\Omega'^k  = \frac{8\epsilon}{\phi_0 r}\int\limits^{+\infty}_{0}df\ \frac{\pi f^2 r^2}{4\phi_0\alpha_0^2}|\delta\tilde\phi(f)|^2 \nonumber\\
&= \mathcal{N}^2(\beta_0,\mu)\frac{2\epsilon \alpha_0^2 M^2}{\pi\phi_0^2 r \tau}. \label{17}
\end{align}
Here the energy-flux $\mathcal{T}^{0k}=(16\pi\alpha_0^2\phi_0)^{-1}\delta\dot\phi^2(t)\hat\Omega '^k$, and the Fourier transform of the scalar field in the zero-frequency limit is $\delta\tilde\phi(f)=\delta\phi_{\text H}/(2\pi i f)$ for $|f|<f_c$, with $\delta\phi_{\text H}$ given by Eq.~(\ref{13}). The coefficient $\epsilon$ comes from the angular part of the integral and as a result of the asymmetric distribution of the scalar field profile. 

\begin{figure}[t]
\centering
\includegraphics[width=95mm]{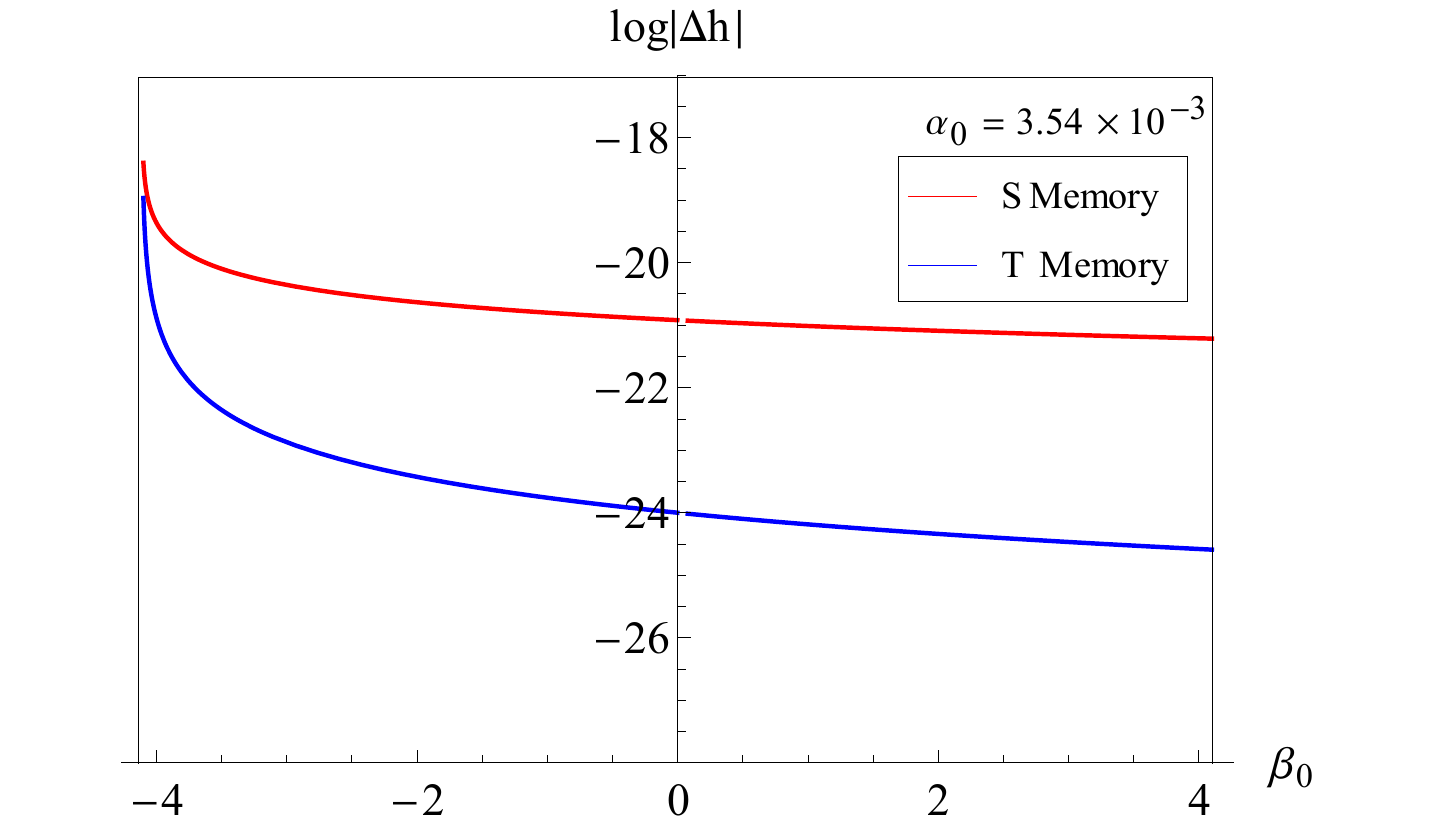}
\caption{Scales of T memory and S memory from gravitational collapse of an $M=10M_\odot$, $R=100M_\odot$,$\epsilon=0.1$ and $r=10\text{kpc}$ Newtonian star.
}
\label{fig2}
\end{figure}

In Fig.~\ref{fig2}, we plot  amplitudes of the  T  and S memory from Eq.~(\ref{14}) and (\ref{17}). In the Brans-Dicke limit, T memory is lower than S memory by about three orders of magnitude. However, since $\Delta h^{\text S}\propto \mathcal{N}(\mu,\beta_0)$ while $\Delta h^{\text T} \propto \mathcal{N}^2(\mu,\beta_0)$, T memory becomes comparable to S memory near $\beta_{\text {crit}}$ where the scalar field profile is significantly magnified by the spontaneous scalarization.

{\noindent {\it Detection Strategies.--}} Since T and S memories contribute to the tensor and scalar components of gravitational wave, respectively, and mixed in observed data, we need a mode separation method~\cite{Hayama} to detect each component. Because the S memory is always larger than T memory in gravitational collapse processes as shown in Fig.~\ref{fig2}, and the existence of non-tensor polarized gravitational wave is a strong evidence for modification to Einstein's theory \cite{Eardley}, we focus on the S memory hereafter. Notice that different gravitational wave detectors on various locations have distinct responses to the three polarizations in Eq.~(\ref{6}), hence it is possible to find linear combinations of the outputs from three or more detectors which only respond to the scalar mode. 

\begin{figure}[t]
\centering
\includegraphics[width=90mm]{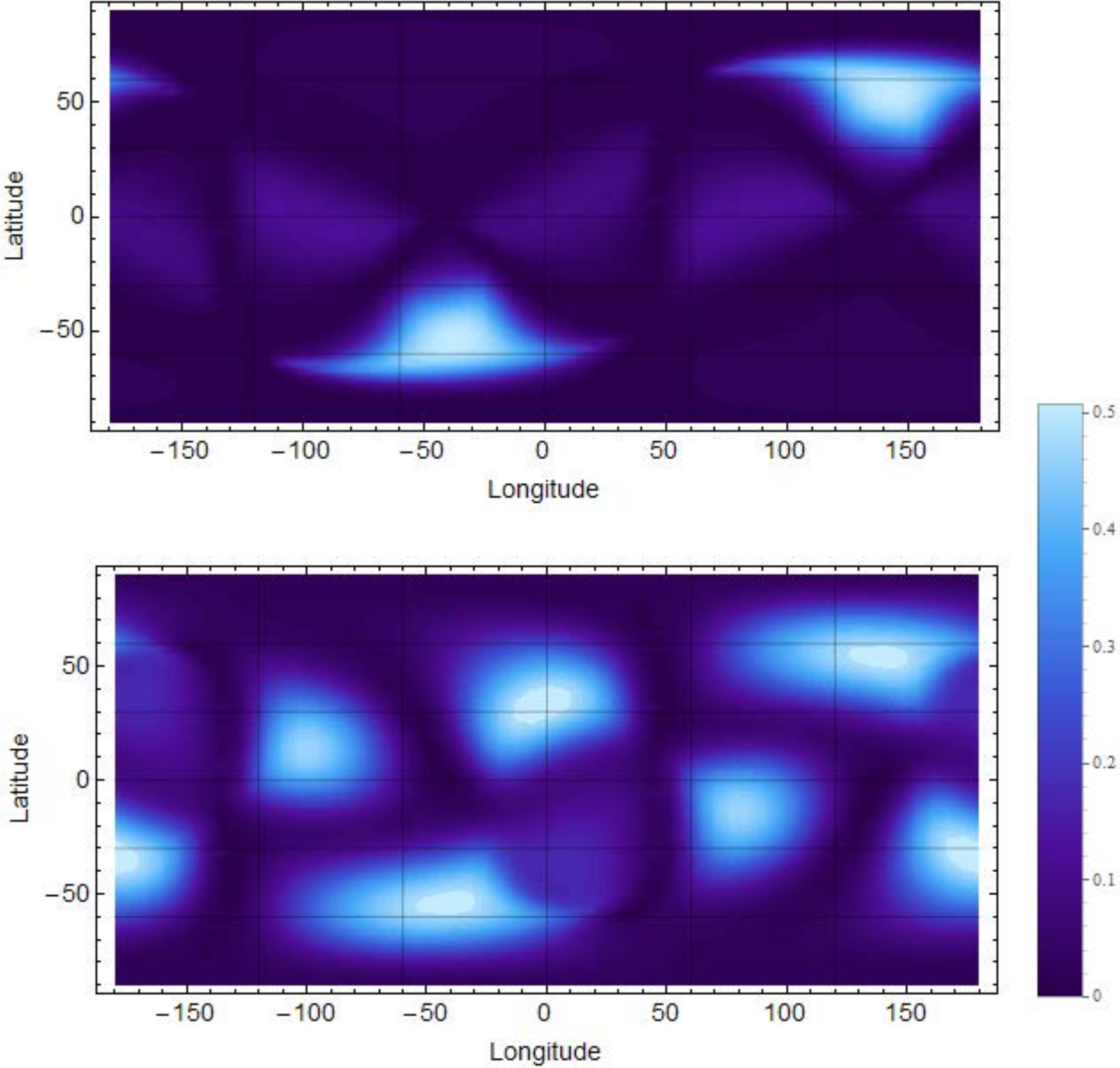}
\caption{The effective angular pattern function $\mathcal F_3(\hat \Omega)$ for network H-L-V (upper panel) and $\mathcal F_4(\hat \Omega)$ for network H-L-V-K (lower panel) as a function of $\hat \Omega$, the direction of the source. The $x$ axis is the longitude as observed on the earth and $y$ axis the latitude.
}
\label{fig3}
\end{figure}

For a network of $N$ detectors, the combined filtered output $W_N$ can be written as $W_N=\vec \alpha\cdot\vec w$, with $\vec \alpha=(\alpha_1,...,\alpha_N)$ the combination coefficients and $\vec w=(w_1,...,w_N)$ the match-filtered outputs of each detector \cite{Thorne}. The signal-to-noise ratio (SNR) of the combined filtered output is defined as $\rho=\text{E}(W_N)/\text{Var}(W_N)^{\frac{1}{2}}$. In order to optimize SNR as well as to make it insensitive to $+$ and $\times$ polarization modes, we should choose
\begin{align}
\vec \alpha=\vec F_\circ-\frac{(\vec F_\times\cdot\vec F_\times)(\vec F_+\cdot\vec F_\circ)-(\vec F_+\cdot\vec F_\times)(\vec F_\times\cdot\vec F_\circ)}{(\vec F_+\cdot\vec F_+)(\vec F_\times\cdot\vec F_\times)-(\vec F_+\cdot\vec F_\times)^2}\vec F_+\nonumber\\
 - \frac{(\vec F_+\cdot\vec F_+)(\vec F_\times\cdot\vec F_\circ)-(\vec F_+\cdot\vec F_\times)(\vec F_+\cdot\vec F_\circ)}{(\vec F_+\cdot\vec F_+)(\vec F_\times\cdot\vec F_\times)-(\vec F_+\cdot\vec F_\times)^2}\vec F_\times ,\nonumber
\end{align}
where $\vec F_P=\left(F^1_P(\hat\Omega),...,F^N_P(\hat\Omega)\right)$, and $F_P^n(\hat\Omega)$ is the angular pattern function of detector $n=1,...,N$ for polarization $P=\circ,+,\times$. The explicit expressions of these functions can be found in \cite{Hayama}. The maximized SNR for the detection of S memory in Eq.~(\ref{14}) is then given by
\begin{align}
\rho=\mathcal{F}^{1/2}_N(\hat\Omega)\frac{2\alpha_0^2M \mathcal{N}(\beta_0,\mu)} {\pi\phi_0 r}\left[\int_0^{f_c}df\ \frac{1}{ f^2 S_n(f)}\right]^\frac{1}{2}. \label{18}
\end{align}
Here for simplicity we suppose the $N$ detectors have approximately the same noise spectral density $S_n(f)$ and $f_c=1/\tau$ is the cut-off frequency of memory as explained above. We describe the dependence of the SNR on the direction of the source by introducing the $N$-detector effective angular pattern function $\mathcal{F}_N(\hat\Omega)$ and is given by
\begin{align}
\mathcal{F}_N(\hat\Omega)=[\vec\alpha\cdot\vec F_\circ(\hat\Omega)]^2/\vec\alpha^2.
\end{align}
We should notice that $\mathcal{F}_N$ is non-zero only for $N\geq3$. We plot $\mathcal F_3(\hat\Omega)$ and $\mathcal F_4(\hat\Omega)$ for network H-L-V and H-L-V-K respectively in Fig.~\ref{fig3}, where H, L, V and K stand for LIGO-Hanford, LIGO-Livingston, Virgo and KAGRA, respectively. For  $\mathcal F_3(\hat\Omega)$, the peak value and the angularly averaged value are 0.485 and 0.087. For $\mathcal F_4(\hat\Omega)$, these values are 0.511 and 0.240. It means that although the inclusion of a fourth detector does not significantly improve the maximal SNR, it does improve substantially the sky coverage of the network.

\begin{figure}[t]
\centering
\includegraphics[width=90mm]{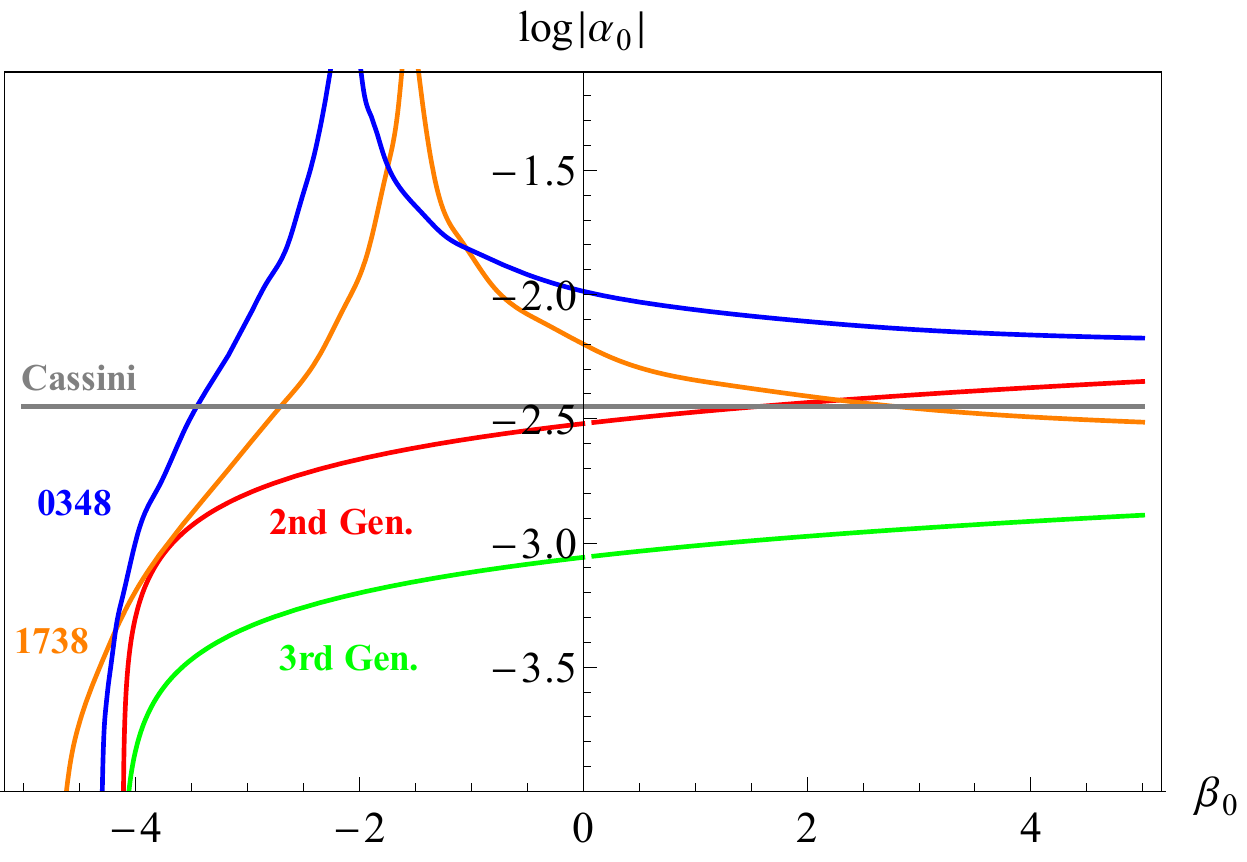}
\caption{Discoverable curves of S memory from a collapsing star with $M=10M_\odot$ $R=100M_\odot$ and $r=10 \text{kpc}$ for 2nd generation detectors (Red) and 3rd generation detectors (Green). The current constraints on the model parameters are from the Cassini Mission (Grey), PSR J1738+0333 (Orange) and PSR J0348+0432 (Blue).}
\label{fig4}
\end{figure}

We next consider the detectability of S memory from our analytic model with $M=10M_\odot$, $R=100M_\odot$ and $r=10 \text{kpc}$. In Eq.~(\ref{18}), we take the threshold SNR to be 10, the effective angular pattern function to be the peak value of network H-L-V-K. We use the design noise spectrum of Advanced LIGO to compute the SNR for second generation detectors and the proposed Einstein Telescope for third generation \cite{Sathyaprakash}. The detectable region of model parameters are shown in Fig.~\ref{fig4}, where we also present current constraints from the solar system  (the Cassini mission) and from pulsar timing (PSR J1738+0333, PSR J0348+0432)~\cite{Berti}. From the figure, the discoverable curves with $\text{SNR} = 1 0$ surpass the current constraints. The gravitational collapse rate is commonly thought to be as low as $\sim$1--3 events per $100$ years in $r<10\text{kpc}$. However, we need to point out that this rate, which is deduced from the SNe rate, is underestimated since more massive stars tend to collapse into black holes directly with no supernova explosion \cite{Fryer}. Besides, such a phenomenon, which, once detected, will provide definitive evidence for the need to modify GR, should not be omitted by future searches in gravitational wave detector data. Hence we propose to add a new search pipeline for the S memory to the upcoming global gravitational wave detector network.

 {\it Discussions.--} In this letter, we have discussed how extra terms in the actions of scalar-tensor theories of gravity and the particular property of black holes in such theories give birth to two new types of gravitational memory, and how these effects can be used as a test of modifications to GR. Another important class of modified gravity is theories with higher order curvature terms, such as Gauss-Bonet theory and Chern-Simons theory \cite{Clifton, Berti}. We expect: (i) the $h_{\mu\nu}^3$ terms in the actions of these theories to be  distinct from GR and hence lead to modifications to the Christodoulou memory, (ii) scalar radiation will continue to cause the T memory, and (iii) since black holes have hair in these theories, the S memory will differ from scalar-tensor theories.  We leave the details for further research.

\noindent {\it Acknowledgements.--} We thank Yanbei Chen for discussions and comments on the manuscript.  Research of SMD and AN was supported by the National Science Foundation, Grant PHY-1404569, PHY-1055103 and the Brinson Foundation.  Research of AN was also supported by the Japan Society for the Promotion of Science and the H2020-MSCA-RISE- 2015 Grant No. StronGrHEP-690904.

\end{document}